\documentclass[useAMS,usenatbib]{mn2e}
\usepackage{amssymb,graphicx,natbib,longtable,hyperref}
\usepackage{breakurl}

\bibpunct{(}{)}{;}{a}{}{,} 

\usepackage[british]{babel}

\title[Hard phase lags in GRO J1744--28]{Discovery of hard phase lags in the pulsed emission of GRO J1744--28}

\author[A. D'A\`i et al.]{A.~D'A\`i$^{1}$ \thanks{antonino.dai@ifc.inaf.it},
L. Burderi$^{2}$, 
T. Di Salvo$^{3}$,
R. Iaria$^{3}$,
F. Pintore$^{4}$,
A. Riggio$^{2}$,
A. Sanna$^{2}$
\and 
\vspace{6pt}\\
$^{1}$ INAF/IASF Palermo, via Ugo La Malfa 153, I-90146 Palermo, Italy\\
$^{2}$ Dipartimento di Fisica, Universit\`a degli Studi di Cagliari, SP Monserrato-Sestu KM 0.7, I-09042 Monserrato, Italy\\
$^{3}$ Dipartimento di Fisica e Chimica, Universit\`a degli Studi di Palermo, via Archirafi 36, I-90123 Palermo, Italy\\
$^{4}$ INAF - IASF Milano, Via E. Bassini 15, I-20133 Milano, Italy\\
}
\date{Received 2016 March 31; accepted 2016 June 1}

\begin{document}

\pagerange{\pageref{firstpage}--\pageref{lastpage}} \pubyear{2016}

\maketitle

\begin{abstract}
We report on the discovery and energy dependence of hard phase lags in
the  2.14 Hz  pulsed  profiles of  GRO J1744--28.   We  used data  from
\textit{XMM--Newton}  and   \textit{NuSTAR}.  We  were  able   to  well
constrain the lag spectrum with  respect to the softest (0.3--2.3 keV)
band: the delay  shows increasing lag values reaching  a maximum delay
of  $\sim$\,12\,ms, between  6 and  6.4 keV.  After this  maximum, the
value of the  hard lag drops to $\sim$\,7\,ms, followed  by a recovery
to a  plateau at $\sim$\,9\,ms  for energies above 8  keV.  $NuSTAR$
data  confirm this  trend  up  to 30  keV,  but  the measurements  are
statistically   poorer,  and   therefore,   less  constraining.    The
lag-energy  pattern up  to the  discontinuity is  well described  by a
logarithmic function. Assuming this is  due to a Compton reverberation
mechanismì,    we   derive    a   size    for   the    Compton   cloud
$R_{\rm{cc}}$\,$\sim$\,120  $R_{\rm  g}$,   consistent  with  previous
estimates on the  magnetospheric radius.  In this  scenario, the sharp
discontinuity at  $\sim$\,6.5 keV  appears difficult to  interpret and
suggests the  possible influence  of the  reflected component  in this
energy range.  We  therefore propose the possible  coexistence of both
Compton and  disc reverberation to explain  the scale of the  lags and
its energy dependence.
\end{abstract}


\section{Introduction}
The transient,  X-ray binary  pulsar GRO J1744-28,  also known  as the
bursting  pulsar due  to  the presence  of recurrent  type-II
bursts,  went  in  outburst  in   2014  mid-January,  reached  a  peak
luminosity  in mid-March,  and returned  to quiescence  at the  end of
April.   During  this   outburst,  it  was  observed   by  many  X-ray
observatories:    results   from    spectral   analysis    done   with
\textit{XMM-Newton}    and   \textit{INTEGRAL}    are   reported    in
\citet{dai15},  with \textit{Chandra}  in \citet{degenaar14}  and with
\textit{Chandra} and \textit{NuSTAR}  in \citet{younes15}.  The pulsar
($P_{\rm  spin}$\,=\,467   ms)  orbits  in  a   nearly  circular  path
(eccentricity $<$\,1.1\,$\times$\,10$^{-3}$) with  a period of 11.8337
d, and a projected semi-major axis  of 2.6324 lt-s.  The mass function
(1.3638\,$\times$\,10$^{-4}$ M$_{\odot}$) indicate an evolved low-mass
(M$_{\odot}$\,$\sim$\,0.2\,M$_{\odot}$)  companion star  and a  nearly
face-on inclination angle \citep{finger96}.

The X-ray pulsed profile can be satisfactorily decomposed into the sum
of two sinusoids,  where the first harmonic accounts  for about 3--8\%
of  the overall  harmonic content  \citep{finger96, dai15,  younes15}.
The  profile's  shape  is  energy dependent; the  pulse  amplitude  is
generally  positively  correlated  with   energy,  with  an  amplitude
fraction of $\sim$\,5\%  for energies below 2 keV,  increasing to more
than 20\% at  $\sim$\,10 keV.  The amplitude fraction  has, however, a
drop  in  the energy  range  corresponding  to the  iron  fluorescence
emission, because of the relative  higher contribution to the observed
flux from incoherent emission from the  reflection component
\citep{dai15}.

Calculations on the long-term evolution of the system that account for
the  present-day orbital  parameters \citep{rappaport97},  measures of
the neutron star (NS) spin-up  rate \citep{finger96}, and evidence for
a  propeller state  at  the  end of  its  1996 outburst  \citep{cui97}
indicated  an intermediate  NS  dipole  field of  a  few 10$^{11}$  G.
Detection of a cyclotron resonance feature  between 4 and 5 keV in the
X-ray spectrum confirmed such expectations \citep{dai15, doroshenko15}
and firmly  established GRO J1744--28  as a peculiar  system in between
the class of the highly magnetized (B\,$\gtrsim$\,10$^{12}$ G), young,
X-ray  pulsars and  the class  of  old NSs  with low-mass  companions,
i.e. accreting  millisecond X-ray  pulsars (AMXPs) and  not-pulsating accreting
NSs,  owing to a  presumed  low magnetic  dipole field  (B\,$\lesssim$\,
10$^{9}$ G).

In this  Letter, we focus on  the relation between the  pulse phase and
energy during the persistent (non-bursting) emission of the source. We
show the existence  of energy-dependent hard phase lags,  we study its
dependence and propose a physical scenario  using a mix of Compton and
disc reverberation.

\section{Observation, data reduction, and analysis}
\textit{XMM--Newton} observed GRO J1744-28 on  2014 March 6. Details on
this observation are  reported in \citet{dai15}. For this  work, we use
only data  from the EPIC/pn  instrument, screening from  type-I bursts
and  considering only  the persistent  emission.  We  applied standard
filtering  criteria  and  included  events  only  in  the  RAWX[28:46]
columns. The light curve exposure is  64.1 ks and the averaged rate is
$\sim$\,660  cps.   Events  were   barycentred  with  respect  to  the
Barycentric Dynamical Time   using the \texttt{barycen} tool with
RA\,=\,266.137792   and   Dec.\,=\,--28.740803  \citep{gosling07}.    To
correct the pulse arrival times for  the orbital motion of the pulsar,
we  adopted  the  following  orbital  parameters  ($Fermi$/GBM  pulsar
project\footnote{
  \url{http://gammaray.msfc.nasa.gov/gbm/science/pulsars/lightcurves/groj1744.html}}):
$P_{\rm   orb}$\,=\,11.836397   d,  $a$~sin($i$)\,=\,2.637   lt-s,   and
$T_{\rm \pi/2}$\,=\,56695.6988 MJD.  We then  used an epoch-folding search
(\texttt{efsearch} in \textsc{xronos}) to find the averaged pulse period during
the  persistent  emission.  The  $\chi^2$  highest  peak is  found  in
correspondence with a period of 0.4670450(1) s.  We selected 20 energy
bands  in  the   EPIC/pn  spectrum,  keeping  the   number  of  events
approximately equal in each selection.  The softest band comprises the
0.3--2.3 keV  range, while  the hardest band  comprises the 9.18--12  keV range.
For  each energy-filtered  event file,  we obtained  the corresponding
folded profile using as  folding period $P_{\rm spin}$\,=\,0.4670450 s
and a phase-bin time of $\sim$\,0.5 ms.  We fitted the pulsed profiles
using the best-fitting function:
\begin{equation}
f(x) = C + A_1  {\textrm sin}(2 \pi (x - \Delta \phi_1)) + A_2 {\textrm sin}(4 \pi (x - \Delta \phi_2))
\end{equation}
where $C$,  $A_1$, $A_2$, $\Delta  \phi_1$, and $\Delta \phi_2$  are the
averaged  profile count  rate, the  sinusoidal semi-amplitudes  of the
fundamental  and  first  overtone  and their  phase  differences  with
respect to the best-fitting phase of the first interval, respectively.

\begin{figure}
\centering
\includegraphics[height=\columnwidth, width=0.8\columnwidth,  angle=-90]{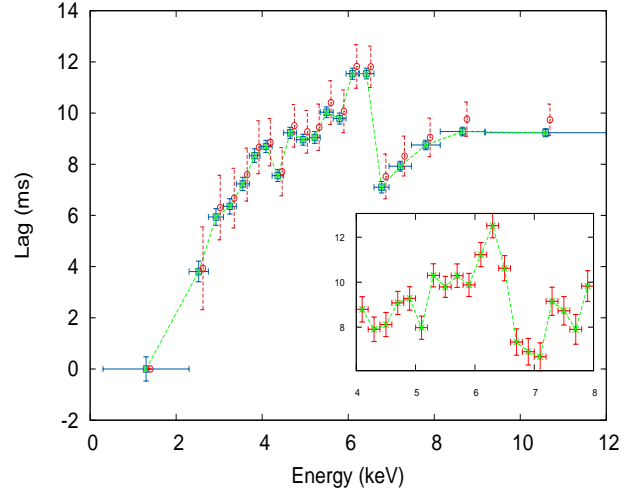}
\caption{Phase-lag dependence with respect  to the softest energy band
  ($E$\,$<$\,2.3  keV).  Blue  open squares:  data from  least-squared
  fits  to   the  folded  profile.    Red  open  circles:   data  from
  cross-correlation  technique,  shifted  of  0.1 keV  in  energy  for
  clarity's  sake.   The  inset  shows the  phase  dependence  in  the
  4.0--8.0 keV  range, adopting a finer  grid (step of 0.2  keV width)
  for the folded profiles with the least-squared fits.}
\label{fig1}
\end{figure}

We obtained  a satisfactory description  of all the profiles,  with an
averaged reduced  $\chi^2$ for  the 20  fits of  1.00 (929  d.o.f. per
fitted  profile)  and  a  standard  deviation of  0.05.   We  show  in
Fig.~\ref{fig1}  the results  for  the $\Delta  \phi_1$  (in units  of
millisecond time difference with respect  to the first energy selected
interval).   Because  we  report  the   phases  with  respect  to  the
best-fitting phase of the first  interval, there is a systematic error
of $\sim$\,1  ms in the $y$-axis,  while error-bars in the  plot show
only the statistical error at 68\% confidence level.  There is a clear
hard-phase  lag that  from the  softest  band rises  up to  12 ms  for
energies  $\sim$\,6  keV.   Above  6.5  keV  the  lag  sharply  drops,
progressively  recovering  to  form   a  plateau  for  energies  above
$\sim$\,8  keV.  Adopting  also  a different  energy  binning (with  a
constant  energy width  of  200 eV)  and repeating  the  steps of  our
analysis, we ensured  that the drop is not an  artefact of the binning
choice (see the inset in Fig.~\ref{fig1}).

We applied  a second procedure to  check the robustness of  the trend,
using    the   cross-correlation    technique.    We   obtained    the
cross-correlation function (CCF) of  each energy-selected profile with
respect to  the first folded  profile. The  resulting CCF had  a clear
bell-shaped function; we obtained the  centre of this function through
a  Gaussian fit  in  a restricted  interval around  the  CCF peak.  To
estimate the error  on the data, we generated for  each profile 1000
simulated  profiles  based on  the  same  statistics of  the  original
profile. We found  again the set of best-fitting centre  values and 
choose  to  associate   as  error  the  standard   deviation  of  this
sample. The final result is shown in Fig.~\ref{fig1} (red open circles
points) together with the results from the first method to allow for a
quick  comparison.  The  trend  is fully  compatible  with  our  first
procedure, the best-fitting lags  are slightly above the corresponding
best-fitting values of  the least-squared lags (still  well within the
error bars) but the associated errors  are systematically larger due to
the   greater   uncertainty   in    estimating   the   peak   of   the
CCF.

We  also analysed  two  \textit{NuSTAR} observations  of GRO  J1744--28
performed on 2014 February 15 (OBS.Id 80002017002, net exposure 24 ks,
rate 185  cps) and on 2014  March 3 (Obs.ID 80002017004,  net exposure
7.1 ks,  rate 570 cps). We  selected a circular region  centred on the
source coordinates of 100 arcsec radius for both observations, and
applied  the same  procedure for  the  analysis of  the EPIC/pn  data,
selecting 12 energy bands.  To improve the statistics, we combined the
light curves  of the  two Focal Plane Modules  (FPMA and
FPMB) to derive  a single folded profile. We obtained  as best folding
periods  0.4670460  s  and  0.4670453  s, for  the  first  and  second
observation, respectively.

\begin{figure}
\centering
\includegraphics[height=\columnwidth, width=0.8\columnwidth,  angle=-90]{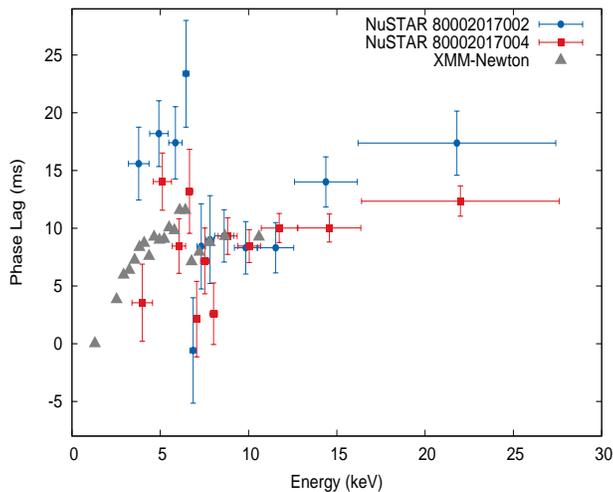}
\caption{Phase-lags for the two $NuSTAR$ observations: in blue and red
  data    from   the    2014-02-15   and    2014-03-03   observations,
  respectively. We  applied a  small shift  of 0.1  keV in  energy for
  clarity's  sake. Grey  filled triangles  show the  $XMM--Newton$ data
  points.}
\label{fig2}
\end{figure}

In Fig.~\ref{fig2} we  show the phase-lag dependence  according to the
least-square  method.    To  allow   a  direct  comparison   with  the
\textit{XMM--Newton}  data,  we  overplotted  them  (grey  triangles),
setting  the  reference phase  of  the  phase lag at  $\sim$\,8.8  keV
channel to be equal to the  value found for the EPIC/pn data set.  The
$NuSTAR$ data show in both  observations qualitatively a trend similar
to the one  shown by EPIC/pn data. Quantitatively, the  lag values are
much  less  constrained  and  they  do  not  allow  us  to  draw  firm
conclusions about possible  differences in the trends  derived for the
two observations,  although there is  a hint  for greater lags  in the
$NuSTAR$  observation of  February, when  the source's  luminosity was
about a third of the March observation.

We  conservatively  limit the  discussion  to  the much  more  clearly
constrained dependence outlined by the EPIC/pn data.

\section*{Discussion}

Energy-dependent soft  phase lags in accreting  X-ray pulsars
have  been  reported  for  sources  belonging  to  the  class  of  the
millisecond         pulsars        \citep{cui98,galloway02,galloway05,
  ibragimov11,falanga12}.   Recently  \citet{miyasaka13}  reported  an
energy  dependent  hard  phase lag for  the  slowly  spinning
($P_{\rm{spin}}$\,=\,12.29  s),  X-ray  pulsar  GS  0834-430,  and  an
energy-modulated   phase-lag  has   been  reported   for  4U   0115+63
\citep{ferrigno11}. The fractional delay for  AMXPs is of the order of
5\% and soft photons lag  harder ones, whereas the hard phase
lag in GS 0834--430 shows a monotonic rise with energy attaining a 30\%
of the pulse period above 50 keV.  GRO J1744--28 energy dependent phase
lag  more closely  resembles the  behaviour  of GS  0834--430, but  its
fractional delay (2.6 per cent of the pulse period) is significantly lower.

Different  interpretations have  been widely  discussed in  literature
about the  origin and  the possible diagnostic  tools offered  by such
measurements \citep{poutanen01}.  A  straight geometric interpretation
of  the lags  due  to  the different  geometrical  paths travelled  by
photons of  different energy bands are  advocated in the case  of some
accreting X-ray binaries and AGNs, where  part of the hard photons are
reflected by  the surface  of the  accretion disk. The  lag is  thus a
measure of  the reverberation distance  between the primary  source of
the  incident  photons and  the  averaged  radius  in the  disc  where
reflection actually  occurs; Fourier-resolved time-lags  have provided
an  additional method  to  connect the  intrinsic  variability of  the
irradiating    flux    with    the    scattered/reflected    component
\citep{uttley14}.

\begin{figure}
\centering
\includegraphics[height=\columnwidth, width=0.8\columnwidth,  angle=-90]{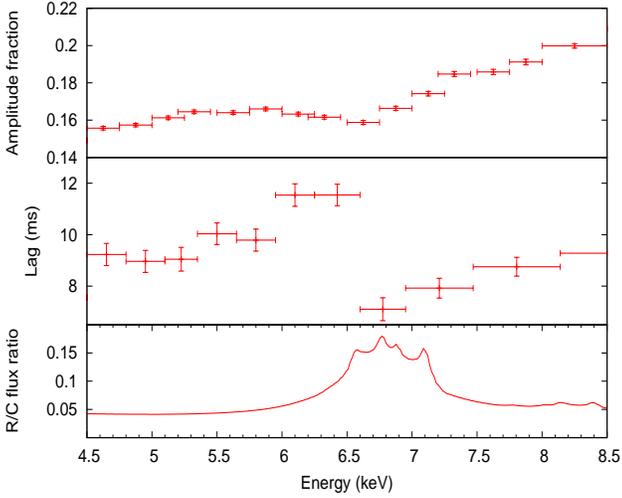}
\caption{Amplitude  fraction  (upper panel)  and  phase-delay
  (middle panel)  of the fundamental harmonic  as function of
  energy in the  4.5--8.5 keV energy range.  Ratio of  the fluxes from
  the reflected and the  irradiating continuum (lower panel),
  according to the spectral decomposition in \citet{dai15}.}
\label{fig:3}
\end{figure}

The non-monotonic lag-energy spectrum of Fig.~\ref{fig1} clearly shows
a smooth rising  trend and a discontinuity in the  iron range, where a
broad   emission   line   is    present   in   the   energy   spectrum
\citep{dai15}.  We   shall  therefore  explore  a   scenario  where  a
combination of  Compton and disc  reverberation might both  operate to
produce the observed lag dependence.  We start making some preliminary
considerations on the  size of its accretion disk. Assuming  an NS mass
of   1.69\,M$_{\odot}$,  a   companion's  mass   of  0.24\,M$_{\odot}$
\citep{rappaport97}   and   the   binary    period   of   11.83   d
\citep{finger96},  we  find  that  the pulsar  Roche  lobe  radius  is
$\sim$\,43  lt-s,  and  the   circularization  radius  $\sim$\,4  lt-s
\citep[Eq. 4.6 and 4.21 in ][]{frankkingraine}, so that the outer disc
radius, which  should lie  in between these  two values,  would easily
accommodate the  overall variation  scale of  the observed  lags.  The
inner disc radius,  if truncated at the  magnetospheric radius, should
instead be  of the  order of 1  lt-ms \citep{dai15}.\\  The reflection
component in the  energy spectrum of GRO J1744--28  gives its strongest
contribution  between  6.6--6.8 keV,  because  of  the presence  of  a
moderately broadened  emission line  from highly ionized  He-like iron
ions  \citep{degenaar14,  dai15, younes15}.   The ratio  $R$  of  the
reflected and the continuum fluxes is  almost constant in the 2--6 keV
range at $\sim$\,5\% and rapidly rises  up to $\sim$\,15\% at the iron
peak (see lower panel of  Fig.\ref{fig:3}).  Presence of this emission
is  also  noticeable  from  the  clear drop  in  the  pulsed  emission
amplitude at this energy, because the reflected component, coming from
at least a distance of 1 lt-ms, has lost coherence with respect to the
non-incident continuum \citep[see upper  panel of fig.~\ref{fig:3} and
  also  Fig.   12 in][]{dai15}.   It  is  therefore plausible  that  a
distortion of the  pulsed profile can be caused  by some reverberation
on the  disk, if the  reflected emission  is seen pulsating  (at least
partially), but  with a  phase difference with  respect to  the direct
emission.   The  energy-dependent  sign  of the  lags  from  the  iron
emission  line depends  on  the distance  where  photons are  actually
emitted, so that  they track the complex, broadened,  line shape, with
the extended  red wing and the  Doppler boosted blue peak  coming from
the  inner radii  (shorter  lags), and  the line  core,  close at  the
rest-frame energy, produced  by the more distant  radii (longer lags),
where  relativistic   and  dynamical   effects  are   strongly  damped
\citep{campana95}.   Qualitatively, the  lag  spectrum,  shown in  the
middle  panel  of  Fig.~\ref{fig:3}, reproduces  these  features:  the
overall scale of  the lag in this energy range  is 3--5 ms, consistent
with the dimensions of the inner parts of the accretion disc. When the
line emission drops,  the lag values at the extreme  wings of the line
are  consistent.  We  also  note  that the  highest  lag  value is  at
$\sim$\,6.4  keV, where  \citet{younes15}  detected possible  emission
from neutral/lowly ionized iron, which may hint also to a contribution
from a neutral and more distant reflector.

When  the energy-lag  spectrum is  considered below  6.5 keV,  the lag
values  rise almost  monotonically. As  a possible  mechanism able  to
produce this monotonic part of the  lag trend, we consider the Compton
reverberation  scenario  \citep{payne80,guilbert82}.   In  this  case,
energetic, relatively hot electrons up-scatter colder photons produced
by  bremsstrahlung,   synchrotron  processes  or  emerging   from  the
thermalized mould at  the base of the accretion column.   If the delay
is produced because  hard photons spend more time  in the Comptonizing
cloud due to repetitive scattering,  we can derive from the dependence
of the  delay on  the photon  energy an  estimate on  the size  of the
cloud,  assuming reasonable  values for  the electron  temperature and
optical depth,  together with  the very  simplifying assumption  of a
uniformly dense and static corona  \citep[but see][for a corona with a
  density gradient]{hua97}.
  
Because of the clear discontinuity at $\sim$\,6.6 keV, we first fitted
the   phase    delays   obtained    from   the    least-squared   fits
(Fig.~\ref{fig1}, blue  open squares), in the  restricted 0.3--6.6 keV
energy range, assuming a logarithmic dependence function, according to
the formula \citep{lightman78}:

\begin{equation} \label{compton}
T_{\rm lag} = \frac{R_{cc}}{c(1+\tau)} \frac{ln(E_h/E_s)}{ln(1+4\Theta(1+4\Theta))},
\end{equation}

where $\Theta=kT_e/mc^2$,  $\tau$ and  $R_{cc}$ are the  optical depth
and the  Compton cloud  size, respectively.  $E_h$  and $E_s$  are the
hard and soft photons energy  ($E_s$\,=\,1.3 keV is a fixed parameter,
being the  central value  of the  first energy band),  and $c$  is the
speed of light.

The logarithmic  function provides  a satisfactory description  of the
delays  up to  6.6 keV  (reduced $\chi^2$  of 1.3  with 15  degrees of
freedom),  with  a  best-fitting  value of  the  constant  before  the
logarithm of 7.0\,$\pm$\,0.1 ms.

There is  small uncertainty  for the cut-off  energy of  the spectrum.
Adopting a  variety of models  and also from  independent observations
and   analysis   of   the   broad-band   spectrum   of   GRO   J1744--28
\citep{dai15,younes15},  the cut-off  energy of  the spectrum  results
$\sim$\,7 keV,  so that  $\Theta \sim 1.3\times10^{-2}$.   The optical
depth is  difficult to  constrain from the  simple observation  of the
slope  of  the  spectrum,  because  the exact  value  depends  on  the
assumptions about the  geometry of the Comptonization region  and on a
variety of concomitant physical processes that are superimposed on the
observed  spectrum  (i.e.   blackbody   emission  from  the  NS  cap,
synchrotron   emission,  and   the  ratio   of  bulk   versus  thermal
Comptonization). Moreover, the scattering cross-section depends on the
direction  of the  photon with  respect to  the magnetic  field lines,
allowing  radiation to  escape more  freely through  the walls  of the
accretion column. We  consider a possible lower  limit of $\tau$\,=\,1
\citep[see  Section\,5.2 in][]{becker07}  and estimate  the region  size
$R_{cc}$ to be of the order  of $\sim$\,0.8 ms-lt (240 km), that would
be consistent with the estimates  on the magnetosphere size as derived
by the  measure of the inferred  NS magnetic dipole and  the dynamical
and  relativistic   broadening  of  the  fluorescence   lines  of  the
reflection component \citep{degenaar14,dai15}.

\begin{figure}
\centering
\includegraphics[height=\columnwidth, width=0.8\columnwidth,  angle=-90]{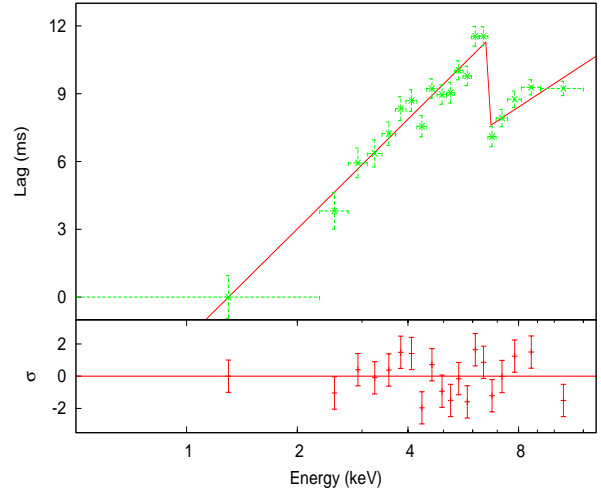}
\caption{Best-fitting piecewise  function (Eq.~\ref{compton},  with a
  break at 6.6  keV) of the EPIC/pn data (Fig.~\ref{fig1})  and residuals in
  units of $\sigma$ (lower panel).}
\label{fig:4}
\end{figure}

We note that after the discontinuity  the trend suggests a return to a
logarithmic dependence as confirmed by the extended $NuSTAR$ coverage,
but with  a different  multiplicative constant (Fig.   \ref{fig1}). In
fact, adopting  a piecewise function, with  f(x)=$k_1 \, log(E_h/Es)$
for $E$\,$<$\,6.6 keV and  $f(x)$=$k_2 \, \textrm{log}(E_h/Es)$ for $E$\,$>$\,6.6
keV,   we   obtained   a  general   satisfactory   fit   ($\chi^2_{\rm
  red}$\,=\,1.47  for  18  dof)  as shown  in  Fig.~\ref{fig:4}.   The
best-fitting   constant   of    this   second   logarithmic   function
($k_2$\,=\,4.5 ms) would point  to physical characteristics similar to
the ones  of the  previously discussed region.  In conclusion,  if the
whole   lag  spectrum   is  interpreted   only  as   due  to   Compton
reverberation, we would  need an unrealistic break in  the iron range,
requiring two Comptonizing separate  regions but with similar physical
parameters.  On  the contrary, it  appears more likely that  the sharp
discontinuity is  determined by  the overimposed  effect of  the disc
reverberation.  In  this case, it remains  to explain why at  the iron
core energy the  overimposed hard lag is not  above the extrapolated
best-fitting function  that fits  the softer 1--6.4  keV lag  range of
Fig.~\ref{fig:4}.  We argue that if the continuum emission is emitted in
a fan-beam  geometry, as possibly  envisaged for pulsars  accreting at
very high $\dot{M}$ \citep{basko76}, and  the GRO J1744--28 is a system
seen  at  low inclination  angle,  with  the magnetic  columns  almost
parallel  to our  line-of-sight, the  optical paths  of the  continuum
photons hitting the disc and directed along our line of sight could be
different. Part  of the continuum  emission that impinges on  the disc
escapes  laterally from  the base  of  the shock  region, whereas  the
Comptonizing  cloud that  produces the  continuous phase-shift  of the
pulsed emission  is along our  line of sight. Although the  details of
this combined  process could  be much  more complex,  we note  that at
least  qualitatively this  sketched  scenario could  explain both  the
scale of the observed lags and  its trend.  We note, that, besides the
Compton reverberation, there  are other ways to  produce a logarithmic
lag-energy  dependence: e.g.  as suggested  by \citet{kotov01}  if the
slope of the  power-law depended on the distance,  with harder indices
produced in  the inner  regions, and  softer in  the outer  regions, a
logarithmic lag-energy spectrum is naturally produced.  Moreover, this
mechanism  coupled with  a reflecting  disk, would  produce a  complex
lag-spectrum,      with     a      dimple     \citep[also      defined
  by][antilags, as they anticipate the softer band]{kotov01}
at the  energies of the fluorescent  iron line very similarly  to what
observed  in  the  GRO  J1744--28.\\   Finally,  we  also  mention  the
possibility that  the lag discontinuity  might be caused by  an abrupt
change  in  the  energy-dependent   scattering  cross-section  at  the
cyclotron resonant energy \citep[as suggested for the accreting pulsar
  4U  0115+63  by][]{ferrigno11},  though the  results  from  spectral
analysis in \citet{dai15} indicated the presence of a moderately broad
absorption feature at lower energy $\sim$\,4.7 keV.

\section{Acknowledgements} 
\small  The High-Energy  Astrophysics  Group  of Palermo  acknowledges
support from the Fondo Finalizzato alla Ricerca (FFR) 2012/13, project
N.      2012-ATE-0390,     founded     by    the     University     of
Palermo.\\  A.\  R.\  gratefully acknowledges  the  Sardinia  Regional
Government  for the  financial support  (P.\ O.\  R.\ Sardegna  F.S.E.
Operational Programme  of the Autonomous Region  of Sardinia, European
Social Fund 2007-2013  - Axis IV Human Resources,  Objective l.3, Line
of  Activity  l.3.1).\\ The  Palermo  and  Cagliari University  groups
acknowledge  financial   contribution  from  the   agreement  ASI-INAF
I/037/12/0.

\bibliographystyle{mn2e}
\bibliography{refs}
\end{document}